\documentclass[aps,prd,showpacs,superscriptaddress,groupedaddress]{revtex4}  
\usepackage{graphicx}  
\usepackage{dcolumn}   
\usepackage{bm}        
\usepackage{amssymb}   
\usepackage{amsmath}   

\usepackage{color}
\usepackage[normalem]{ulem}

\begin{document}


\title{Non-leptonic kaon decays at large $N_c$}

\date{\today}

\author{A.~Donini}
\affiliation{IFIC (CSIC-UVEG), Edificio Institutos Investigaci\'on, 
Apt.\ 22085, E-46071 Valencia, Spain}
\author{P.~Hern\'andez}
\affiliation{IFIC (CSIC-UVEG), Edificio Institutos Investigaci\'on, 
Apt.\ 22085, E-46071 Valencia, Spain}
\author{C.~Pena}
\affiliation{Departamento de F\'{\i}sica Te\'orica and Instituto de F\'{\i}sica Te\'orica UAM-CSIC,
Universidad Aut\'onoma de Madrid, E-28049 Madrid, Spain}
\author{F.~Romero-L\'opez}
\affiliation{IFIC (CSIC-UVEG), Edificio Institutos Investigaci\'on, 
Apt.\ 22085, E-46071 Valencia, Spain}

\preprint{IFIC/16-39}
\preprint{IFT-UAM/CSIC-16-063}
\preprint{FTUAM-16-26}

\begin{abstract}
We study the scaling with the number of colours, $N_c$, of the weak amplitudes mediating kaon mixing and decay. We evaluate the amplitudes of the two
relevant current-current operators  on the lattice for $N_c=3-7$. We conclude that the subleading  $1/N_c$ corrections in
$\hat{B}_K$  are small, but those in the  $K \rightarrow \pi\pi$ amplitudes are large and fully anti-correlated in the $I=0, 2$ isospin channels. We briefly comment on the implications for the $\Delta I=1/2$ rule. 
\end{abstract}
\maketitle


\section{Introduction}

The prediction of flavour violating processes involving kaons remains elusive.  In particular, there is still no satisfactory explanation of the striking $\Delta I=1/2$ rule, nor a reliable prediction of $\epsilon'/\epsilon$. In spite of the spectacular progress in lattice QCD calculations in the past decade, few attempts have been made at these difficult observables, and the systematic uncertainties in the existing results \cite{Bai:2015nea} remain 
too large. On the other hand, a rather precise determination of the $K-\bar{K}$ mixing amplitude
(given by $\hat{B}_K$) has emerged \cite{Aoki:2013ldr,BK}. 

The large $N_c$ limit of QCD \cite{'tHooft:1973jz} has been  invoked in many phenomenological approaches to this problem  (some relevant references are \cite{Buras:2014maa,Pich:1995qp,Peris:2000sw,Hambye:2003cy,Cirigliano:2011ny}). This seems counter-intuitive since the strict large $N_c$ limit of the $\Delta I=1/2$ rule fails completely. The predictions therefore 
rely on significant sub-leading $N_c$ effects, which are however very difficult to predict accurately. As a result,  these approaches typically involve 
further approximations beyond  the strict large-$N_c$ expansion. 

In \cite{Boyle:2012ys}, the results of the most ambitious lattice computation of $K \rightarrow \pi\pi$ to date were presented, and a significant $\Delta I=1/2$ dominance was observed. It was noted that the $\Delta I=1/2$ rule seems to be originating in an approximate cancellation of the two diagrams (color connected and disconnected) 
contributing to the $\Delta I=3/2$ amplitude. Unfortunately it is not possible to isolate these two contributions physically,  so it is not clear what
to extract from this finding. In the large $N_c$ expansion however this is possible since the leading scaling in $N_c$  of the contributions is different.  
The cancellation can therefore be phrased in terms of the sign and size of the $1/N_c$ corrections in the  isospin amplitudes. In fact, it was in the context of phenomenological approaches using the large $N_c$ expansion where the opposite sign of these contributions was first pointed out \cite{Pich:1995qp}. There is however a strong correlation between
the $\Delta I=3/2$ amplitude and $\hat{B}_K$ and therefore this suggest that the same cancellation 
in the former should be affecting the latter, suggesting a value of $\hat B_K$ significantly smaller  than the $N_c\rightarrow \infty$ value.
The role of the $1/N_c$ expansion in the interpretation of the
results in \cite{Boyle:2012ys} is also discussed in the latest update of RBC/UKQCD's results
for the $K\to\pi\pi$ $\Delta I=3/2$ decay amplitude~\cite{Blum:2015ywa}.
A study of the --- related --- issue of deviations from the na\"{\i}ve factorization approximation
to $K\to\pi\pi$ amplitudes can be found in~\cite{Carrasco:2013jda}.

The goal of this paper is to study from first principles the large $N_c$ behaviour of  certain $\Delta S=1$
and $\Delta S=2$ amplitudes. More concretely we consider 
$K$-$\pi$ and $K$-$\bar{K}$ transitions mediated by the four-fermion current-current operators on the lattice varying the number of colours $N_c=3-7$. As it is well known,  these amplitudes fix $\hat{B}_K$ (up to $SU(3)$ flavour breaking effects by quark masses) and, up to chiral corrections, also the $\Delta I=3/2$ contribution to the non-leptonic kaon decay, $K \rightarrow \pi\pi$  \cite{di32}. Furthermore, in the GIM limit of degenerate charm and up quarks,  the $\Delta I=1/2$ contribution to the non-leptonic decays can also be determined from the current-current operator matrix elements, only \cite{Giusti:2004an,Giusti:2006mh}. In fact this is the limit where the cancellation of \cite{Boyle:2012ys} can be more clearly isolated. For this reason,  we will consider only the $SU(4)$-flavour limit $m_c=m_u=m_d=m_s$. We miss in this way the effects of a heavy charm, which were originally argued  to be the origin of the $\Delta I=1/2$ rule \cite{Shifman:1975tn}. 
This fact, however, has not been confirmed by non-perturbative studies \cite{Bai:2015nea,Endress:2014ppa}. 

The paper is organized as follows. In section~\ref{sec:formalism} we introduce our method and set our notation. We present the main results in section~\ref{sec:results} and conclude in \ref{sec:conclusions}.

\section{Formalism}
\label{sec:formalism}

The Operator Product Expansion  allows to represent the weak Hamiltonian that mediates  CP-conserving $\Delta S=1$ transitions by an effective Hamiltonian in terms of four-fermion operators. At the electroweak scale, $\mu \simeq M_W$, we can neglect all quark masses and the weak Hamiltonian takes the simple form:
\begin{gather}
\label{eq:heffs1}
H_{\rm w}^{\Delta S=1} = \int d^4x~\frac{g_{\rm w}^2}{4M_W^2}V_{us}^*V_{ud}\sum_{\sigma=\pm} k^\sigma(\mu) \, \bar{Q}^\sigma(x,\mu)\,,
\end{gather}
where $g_{\rm w}^2=4\sqrt{2}G_{\rm F} M_W^2$. Only two four-quark operators of dimension six can appear with the correct symmetry properties under the flavour 
symmetry group  ${\rm SU}(4)_{\rm L} \times {\rm SU}(4)_{\rm R}$, namely
\begin{gather}
\begin{split}
{\bar Q}^\pm(x,\mu) = Z_Q^\pm(\mu) \, \big(& J_\mu^{su}(x)J_\mu^{ud}(x) \pm J_\mu^{sd}(x)J_\mu^{uu}(x) \\
&~-~[u\leftrightarrow c]\big)\,,
\end{split}
\end{gather}
where $J_\mu$ is the left-handed current, $J_\mu^{\alpha\beta} = (\bar\psi_\alpha\gamma_\mu P_-\psi_\beta)$, 
$P_\pm={1\over 2} (\mathbf{1}\pm\gamma_5)$, and parentheses around quark bilinears indicate that they are traced over spin and colour. 
Eventually, $Z_Q^\pm (\mu)$ is the renormalisation constant of the bare operator $Q^\pm (x)$ computed in some regularisation scheme as, for example, the lattice.
There are other bilinear operators of lower dimensionality that could mix with those above: however, they vanish in the GIM limit  \cite{Giusti:2004an}. 

The operators ${\bar Q}^\sigma(\mu)$ are renormalised at a scale $\mu$ in some renormalisation scheme, being their $\mu$-dependence exactly cancelled by that of the Wilson coefficients $k^\sigma(\mu)$. It is common practice to define renormalisation group invariant (RGI) operators, which are defined by cancelling their perturbative $\mu$-dependence, as derived from the Callan-Symanzik equations,
\begin{eqnarray}
\hat{Q}^\sigma \equiv \hat{c}^\sigma(\mu) {\bar Q}^\sigma(\mu), 
\end{eqnarray}
with 
\begin{eqnarray}
{\hat c}^\sigma(\mu)\equiv \left({N_c \over 3} {g^2(\mu)\over 4 \pi} \right)^{\kern-0.2em-{\gamma^\sigma_0\over 2 b_0}}
\kern-1.6em\exp\left\{ -\kern-0.3em\int_0^{g(\mu)}\kern-1.4em{\rm d}g
\left[{\gamma^\sigma(g)\over \beta(g)} - {\gamma^\sigma_0\over b_0 \, g}\right]\right\},
\label{eq:c}
\end{eqnarray}
where $g(\mu)$ is the running coupling and $\beta(g)=-g^3 \sum_n b_n g^{2 n}$, $\gamma^\sigma(g) = -g^2 \sum_n \gamma^\sigma_n g^{2 n}$ are the $\beta$-function and the anomalous dimension, respectively. The one- and two-loop coefficients of the $\beta$-function, and the one-loop coefficient
of the anomalous dimensions, are renormalisation scheme-independent. Their values for the theory with $N_f$ flavours are \cite{b0,b1}
\begin{eqnarray}
b_0 &=& \frac{1}{(4\pi)^2}\left[\frac{11}{3}N_c-\frac{2}{3}N_f\right]\,,\\
b_1 &=& \frac{1}{(4\pi)^4}\left[\frac{34}{3}N_c^2-\left(\frac{13}{3}N_c-\frac{1}{N_c}\right)N_f\right]\,,
\end{eqnarray}
and for the operators $Q^\pm$ \cite{load}
\begin{gather}
\gamma_0^\pm = \frac{1}{(4\pi)^2}\left[\pm 6 - \frac{6}{N_c}\right]\,.
\end{gather}
The normalisation of $\hat{c}^\sigma(\mu)$ coincides with the most popular one for $N_c=3$, whilst using the 't Hooft coupling $\lambda = N_c g^2 (\mu)$
in the first factor instead of the usual coupling, so that the large $N_c$ limit  is well defined.  

Defining similarly an RGI Wilson coefficient 
\begin{eqnarray}
\label{eq:WilsonRGI}
\hat{k}^\sigma \equiv {k^\sigma(\mu)\over \hat{c}^\sigma(\mu)},
\end{eqnarray}
we can rewrite the Hamiltonian in terms of RGI quantities, which no longer depend on the scale, so we can write
\begin{gather}
\hat{k}^\sigma \, \hat{Q}^\sigma  = \left[ { k^\sigma(M_W) \over \hat{c}^\sigma(M_W) } \right] \, \left[ {\hat c}^\sigma(\mu)\, \bar{Q}^\sigma (\mu) \right]
= k^\sigma(M_W) \, U^\sigma(\mu,M_W) \, \bar{Q}^\sigma (\mu)
\end{gather}
where $\mu$ is a convenient renormalisation scale for the non-perturbative computation of matrix elements of $Q^\pm$, which will be later set to the inverse lattice scale $a^{-1}$. The factor $U^\sigma(\mu, M_W) = {\hat c}^\sigma(\mu)/ {\hat c}^\sigma(M_W)$, therefore, measures the running of the renormalised operator between the scales $\mu$ and $M_W$.
Ideally one would like to evaluate this factor non-perturbatively, as has been done for $N_c=3$ \cite{nprunning}, but this is beyond the scope of this paper. We will instead use 
the perturbative results at two loops in the RI scheme \cite{nload}
to evaluate the $\hat{c}^\sigma(\mu)$ factors. This implies relying on perturbation theory at scales
above $\mu=a^{-1} \sim 2~{\rm GeV}$.

Our goal is to compute the $K\to\pi$ amplitudes mediated by $H_{\rm w}^{\Delta S=1}$.
The hadronic contribution is encoded in the ratios of three- and two-point functions
\begin{eqnarray}
\hat{R}^\pm \equiv \frac{\langle\pi|\hat{Q}^\pm|K\rangle}{f_K f_\pi m_Km_\pi}
= \hat{c}^\pm(\mu) Z_R^\pm(\mu) R^\pm \,,
\label{eq:ratio}
\end{eqnarray}
where $Z_R^\pm(\mu)$ are the renormalisation factors for the ratios and $R^\pm$ is the ratio of matrix elements of bare operators. 
In the $SU(3)$ limit $m_s=m_d=m_u$, from $R^+$ we can determine $\hat{B}_K$ as
\begin{eqnarray}
\hat{B}_K = {3 \over 4} \hat{R}^+ .
\end{eqnarray}
Concerning $K\rightarrow \pi\pi$ decays, the two very different isospin amplitudes
\begin{eqnarray}
i A_I e^{i \delta_I} \equiv \langle (\pi\pi)_I | H_W |K_0\rangle,\;\;\; I=0,2
\end{eqnarray}
can be related in chiral perturbation theory, and in the GIM limit, to the $K\to\pi$
amplitudes $A^\pm  \equiv \hat{k}^\pm \hat{R}^\pm$ \cite{Giusti:2004an}: 
\begin{eqnarray}
{A_0 \over A_2} = {1 \over \sqrt{2}} \left({1\over 2} + {3 \over 2} {A^-\over A^+}\right).
\end{eqnarray} 
The $\Delta I=1/2$ rule, {\em i.e.} the large enhancement of the ratio $|A_0/A_2| \sim 22$, is therefore related in this limit to the ratio of the amplitudes
$A^-/A^+$. 

At this point, it is necessary to comment on the chiral corrections. The relation between the $K-{\bar K}$ and $K\rightarrow (\pi\pi)|_{I = 2}$ amplitudes is well known to break down away from the chiral limit for the physical case $m_s \gg m_{u,d}$, since the chiral logarithmic corrections are much larger for the former amplitude \cite{di32}. On the other hand, this is not the case in the $SU(3)$ limit $m_s=m_u=m_d$, where the chiral logs are the same for both amplitudes both in the full as in the quenched case \cite{Golterman:1997wb}.  The following relation holds up to one loop in ChPT in the leading-log approximation:
\begin{eqnarray}
\left.{\langle \pi^+ \pi^0| H_W | K\rangle \over m_K^2 -m_\pi^2}\right|_{ m_s = m_d} = {i F\over \sqrt{2}} A^+ G_F V_{ud} V_{us}^*, 
\end{eqnarray}
where $F$ is the decay constant in the chiral limit and $A^+$ contains one loop corrections.
This shows that, in this approximation,  the $1/N_c$ corrections in the physical amplitude are
fixed\footnote{
It has been argued that higher-order ChiPT effects may have an important impact on
$K\to\pi\pi$ amplitudes at the same order in $1/N_c$. Some relevant references are~\cite{Truong:1987hn,Isgur:1989js,Kambor:1991ah,Pallante:2000hk}.
}
by those in $A^+$.
At the same order in ChPT, we can relate the amplitudes for both choices of quark masses:
\begin{eqnarray}
 \langle \pi^+\pi^0 | H_W | K^+\rangle_{m_\pi \rightarrow 0} &=&  m_K^2 \left.{ \langle \pi^+\pi^0 | H_W | K^+\rangle \over m_K^2 -m_\pi^2 }\right|_{m_s= m_d} \left(1 + {9\over 4} {m_K^2 \over (4 \pi F)^2 } \log {m_K^2 \over (4 \pi F)^2} \right).
\end{eqnarray}
The chiral log term gives an additional {\it negative}  $1/N_c$ contribution to the amplitude at the physical point with respect to that in the degenerate case. 
Another important point to note is   that, in the GIM limit, the chiral logs have been shown to be fully anticorrelated in $A^\pm$ \cite{Hernandez:2006kz} and therefore an extrapolation to the chiral limit using chiral perturbation theory will not change the anticorrelation found at larger masses. Unfortunately the computation of chiral logs  in $K\rightarrow (\pi\pi)_{I=0}$ in the GIM limit is not available, although it is likely that the same anticorrelation holds also there.

\section{Results}
\label{sec:results}

We compute the ratios $\hat R^\pm$ on the lattice from the ratio of correlation functions
\begin{eqnarray}
\label{eq:bareratios}
R^\pm  = \kern-1.0em
\lim_{ \substack{z_0-x_0\to\infty \\ y_0-z_0\to \infty}}
\frac{\sum_{{\mathbf x},{\mathbf y}}\langle P^{du}(y) Q^\pm(z) P^{us}(x)\rangle}
{\sum_{{\mathbf x,\mathbf y}}\langle P^{du}(y)A_0^{ud}(z)\rangle \langle A_0^{su}(z) P^{us}(x)\rangle}\,,
\end{eqnarray}
where $P^{ab}(x)=\bar{\psi}^a(x) \gamma_5\psi^b(x)$, and $A^{ab}_0(x)=Z_{\rm A}\bar{\psi}^a(x) \gamma_0 \gamma_5\psi^b(x)$. 
The renormalised ratios $\hat R^\pm$ have been computed in $SU(N_c)$ for $N_c=3-7$ and in the quenched approximation. 
Note that the latter does not modify the leading large $N_c$ result, but it can modify the first subleading $1/N_c$ corrections. 
We have implemented the required correlation functions in the source code first developed in \cite{DelDebbio:2008zf} and further optimized in \cite{pica}. 
The number of colours and the lattice size are given in the first two columns of Table~\ref{tab:sim}. The spatial volume, $L/a = 16$, is kept fixed in all simulations. 
On the other hand, $T/a = 48$ for $N_c = 3, 4, 5$ and $T/a = 32$ for $N_c = 6,7$.
Following \cite{Bali:2013kia} the bare coupling, $\beta= 2 N_c/g_0^2$,  is tuned with $N_c$ in such a way that the 
string tension remains constant $a \sqrt{\sigma} \simeq 0.2093$; this results in $a \simeq 0.093~{\rm fm}$ with $\sigma=1~{\rm GeV/fm}$. 
The bare 't Hooft coupling $\lambda$ is found to be well described by the following scaling
\begin{eqnarray}
\lambda = N_c g^2_0 = 2.775(3) +{1.90(3)\over N_c^2}.
\end{eqnarray}
The coupling $\beta$ as a function of $N_c$ is given in the third column of Table~\ref{tab:sim}.
In order to preserve the multiplicative renormalisation of $Q^\pm$, while avoiding the high computational cost of a simulation with exactly chiral lattice fermions, 
we use a Wilson twisted-mass fermion regularisation \cite{tmqcd}. (For the gauge sector we employ the standard plaquette action.)
This allows to devise a formulation of valence quarks that not only
preserves good renormalisation properties, but also prevents the appearance of linear cutoff effects
in $a$ \cite{Frezzotti:2004wz}.
The full-twist condition amounts to having a vanishing current quark mass $m_{\rm PCAC}$ from the axial
Takahashi-Ward identity in so-called twisted quark field variables. The value of $am_{\rm PCAC}$ in our simulations is given in the fourth column of Table~\ref{tab:sim}, 
where we can see that the full-twist condition $am_{\rm PCAC} = 0$, expected from an accurate tuning of the Wilson
critical mass (which we again take from \cite{Bali:2013kia}), is satisfied to a varying degree of
accuracy; the deviations present are however irrelevant within the precision of our results.
The bare quark mass is chosen to provide a pseudoscalar mass not far from the physical kaon mass in all cases (see the fifth column of Table~\ref{tab:sim}).
Eventually, our results for the bare ratios $R^\pm$ defined in eq.~(\ref{eq:bareratios}), computed  in the $SU(3)$ limit, are shown in the last two columns of the table. 

 \begin{table}[!th]
\begin{center}
\begin{tabular}{clcr@{\hspace{0mm}}c@{\hspace{0mm}}lcll}
\hline\hline
$N_c$ & 
$T/a$ &
~~~~$\beta$ &
\multicolumn{3}{c}{$a m_{\rm\scriptscriptstyle PCAC}$} &  $a m_{\rm PS} 
$ & ~~$R^+_{\rm bare}$ & ~~$R^-_{\rm bare}$    \\[0.5ex] 
\hline\\[-2.0ex]
3 & $48$ & 6.0175  & -0&.&002(14)  & 0.2718(61) & 0.774(21) & 1.218(31) \\
4 & $48$ & 11.028  & -0&.&0015(11) & 0.2637(39) & 0.783(15) & 1.198(19) \\
5 & $48$ & 17.535  &  0&.&0028(9)  & 0.2655(31) & 0.839(8)  & 1.145(12) \\
6 & $32$ & 25.452  &  0&.&0013(7)  & 0.2676(28) & 0.871(6)  & 1.125(7)  \\
7 & $32$ & 34.8343 & -0&.&0034(6)  & 0.2819(19) & 0.880(5)  & 1.122(5)  \\
 \hline
\hline\end{tabular}
\caption{Lattice simulation results. Lattice sizes are $(L/a)^3 \times (T/a)$,
with $L/a=16$ throughout. The twisted bare mass
is fixed to $a \mu=0.02$. The lattice spacing is fixed by the string tension through
$a \sqrt{\sigma} \simeq 0.2093$ \cite{Bali:2013kia}. $m_{\rm PCAC}$ is the current mass obtained
from the axial Takahashi-Ward identity in twisted quark field variables. $m_{\rm PS}$
is the kaon and pion mass in our $m_u=m_d=m_s$ limit. $R^\pm$ are our results for the bare ratios given in eq.~(\ref{eq:bareratios}).}
\label{tab:sim}
\end{center}
\end{table}

In Table~\ref{tab:renorm}  we show the various renormalisation constants and RG running factors needed to compute the renormalised amplitudes $\hat B_K$ and $A^\pm$
as a function of the number of colours. First of all, in order to get the renormalised ratios $\hat R^\pm$ from the bare ones computed on the lattice, 
we have used the known one-loop lattice renormalisation constants in the RI scheme of ref.~\cite{etmpert}.
Note that, due to the breaking of chiral symmetry in the adopted regularisation, the axial current requires
a finite, $N_c$-dependent, renormalisation constant $Z_A$, that has to be included in the factors $Z_R^\pm$ in eq.~(\ref{eq:ratio}).
This has also been taken from ref.~\cite{etmpert}. The values of $Z^\pm (a^{-1})$ are given in the rightmost column of Table~\ref{tab:renorm}.
The values of the normalisation coefficients $\hat c^\pm (a^{-1})$ and of the running of the renormalised operators from the scale of lattice computations, $\mu = a^{-1}$, 
to the scale of the effective theory, $M_W$, computed using perturbative results at two-loops in the RI scheme \cite{nload},
 are given in the fifth and fourth columns of Table~\ref{tab:renorm}, respectively.
In the evaluation of the $\hat{c}^\sigma(\mu)$ factors we have used the large $N_c$ scaling of the $\Lambda$ parameter found in 
ref.~\cite{Allton:2008ty},
\begin{eqnarray}
{\Lambda_{\overline{MS}}\over \sqrt{\sigma}} = 0.503(2)(40) +{0.33(3)(3)\over N_c^2}.
\label{eq:lambdamsbar}
\end{eqnarray}
Eventually, the Wilson coefficients $k^\pm (M_W)$, also computed following ref.~\cite{nload}, are given in the third column of Table~\ref{tab:renorm}, while their RGI counterparts $\hat{k}^\pm$, defined in eq.~(\ref{eq:WilsonRGI}), are given in the second column.

\begin{table}[!t]
\begin{center}
\begin{tabular}{c@{\hspace{5mm}}ccccc}
\hline\hline\\[-2.0ex]
$N_c$ &
$\hat{k}^+$ & $k^+(M_W)$ & $U^+(a^{-1},M_W)$ & $\hat{c}^+(a^{-1})$ & $Z^+(a^{-1})$ \\[0.3ex] 
\hline\\[-2.0ex]
 3 & 0.642 & 1.030 & 0.875 & 1.404 & 0.983 \\
 4 & 0.658 & 1.025 & 0.895 & 1.394 & 0.988 \\
 5 & 0.679 & 1.021 & 0.910 & 1.368 & 0.991 \\
 6 & 0.700 & 1.018 & 0.921 & 1.340 & 0.994 \\
 7 & 0.719 & 1.016 & 0.930 & 1.315 & 0.996 \\
\hline\hline
\end{tabular}

\vspace{5mm}

\begin{tabular}{c@{\hspace{5mm}}ccccc}
\hline\hline\\[-2.0ex]
$N_c$ &
$\hat{k}^-$ & $k^-(M_W)$ & $U^-(a^{-1},M_W)$ & $\hat{c}^-(a^{-1})$ & $Z^-(a^{-1})$ \\[0.3ex] 
\hline\\[-2.0ex]
 3 & 2.398 & 0.940 & 1.319 & 0.517 & 1.059 \\
 4 & 1.998 & 0.958 & 1.210 & 0.580 & 1.043 \\
 5 & 1.780 & 0.968 & 1.156 & 0.620 & 1.035 \\
 6 & 1.643 & 0.974 & 1.124 & 0.666 & 1.030 \\
 7 & 1.550 & 0.978 & 1.103 & 0.696 & 1.026 \\
\hline\hline
\end{tabular}
\caption{Perturbative renormalisation constants and RG running factors.  $Z^\sigma (a^{-1})$ at one-loop have been extracted from \cite{etmpert}, whereas $U^\sigma$ and $k^\sigma$ are computed using the two-loop $\overline{\rm MS}$ coupling (with $\Lambda_{\overline{MS}}$ taken from eq.~(\ref{eq:lambdamsbar}) from ref.~\cite{Allton:2008ty}).}
\label{tab:renorm} 
\end{center}
\end{table}

\begin{figure}[!t]
 \begin{center}
 \includegraphics[scale=0.68]{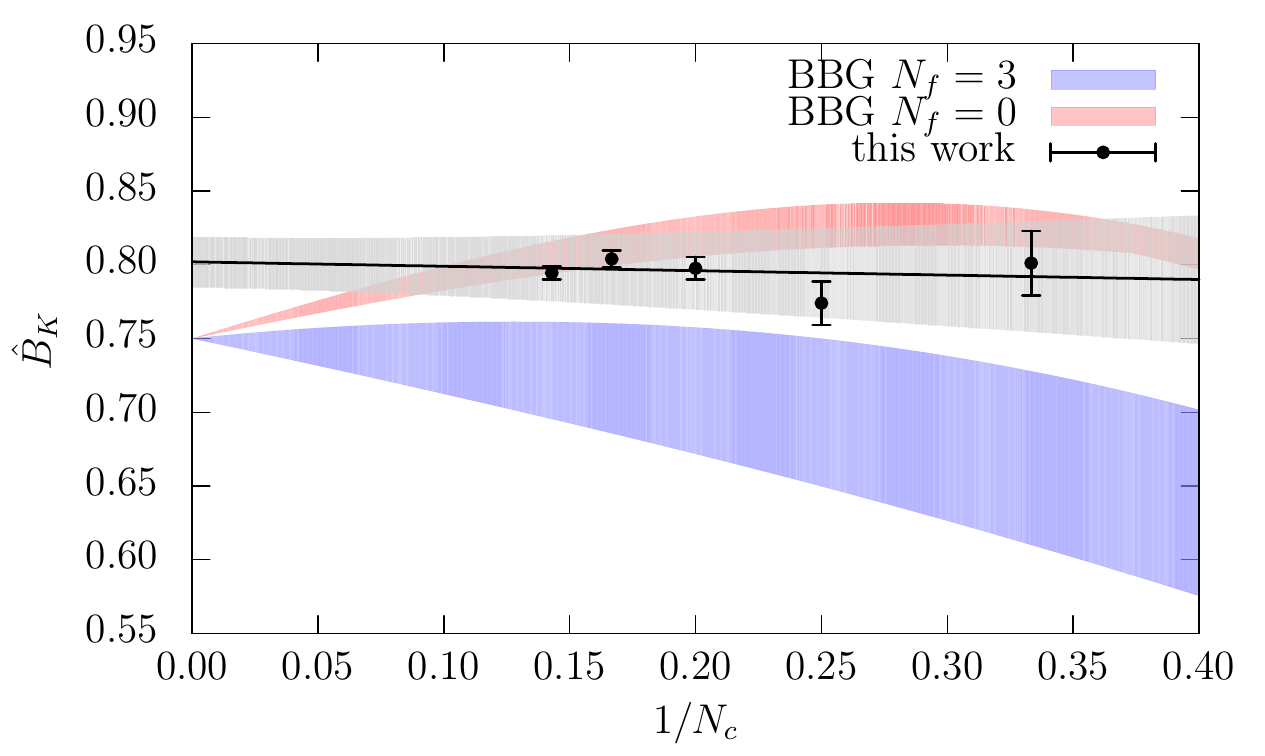}  
\caption{$\hat{B}_K$ versus $1/N_c$.
The grey band (solid line) is a linear fit to our five data points. The red and blue bands use the model prediction of~\cite{Buras:2014maa}.}
\label{fig:bk}
\end{center}
\end{figure}

Our results for $\hat{B}_K$ as a function of $1/N_c$ are shown in Fig.~\ref{fig:bk} together with a linear fit to the data, represented by a solid black line. The grey band 
shows the 1$\sigma$ error on the fit. We compare our results with our own evaluation of the predictions of the phenomenological analysis in ref.~\cite{Buras:2014maa}, 
represented by a light red band for $N_f = 3$ and by a blue band for $N_f = 0$.
For $N_f=3$ we use in the latter the same values for hadronic masses and decay constants as in~\cite{Buras:2014maa}, and obtained the decay constant for $N_c \neq 3$ by rescaling
$F_K=F_K(N_c=3)\sqrt{N_c/3}$. For $N_f=0$ we use as input for the hadronic quantities, including their $N_c$ dependence, the interpolating formulae provided in~\cite{Bali:2013kia},
matched to our measured values of $M_K$. In both cases the band covers the difference between setting the matching scale $M$ in eq.~(62) of~\cite{Buras:2014maa} at $0.6~{\rm GeV}$ and at $1~{\rm GeV}$; for $N_f=0$ it also comprises the uncertainty due to our value of $M_K$ not being constant within errors as a function of $N_c$.
Notice that both theoretical predictions give $\hat B_K = 3/4$ in the $N_c \to \infty$ limit. From Fig.~\ref{fig:bk} we can see that the subleading $1/N_c$ corrections in $\hat{B}_K$ are small (which goes in the direction of the predictions in \cite{Buras:2014maa}, but not those in \cite{Peris:2000sw}, that correspond to the chiral limit).
The parameter of the linear fit to the data are shown in the first two lines of Table~\ref{tab:fits} for a different choice of the data points included in the fit, together with the corresponding
$p$-values. The third line of the same table shows our result for a quadratic fit to the data. We can see that, in this case, the large $N_c$ limit obtained is consistent with the 
theoretical expectation, albeit with large errors. Note that a significant $O(a^2)$ uncertainty for $R^+$ can be expected, 
cf. the $\mathcal{O}(10\%)$ effect for $N_c=3$, $N_f=2$ shown by the data of \cite{Constantinou:2010qv} at a lattice spacing comparable to ours.

\begin{table}[!t]
\begin{center}
\begin{tabular}{c@{\hspace{5mm}}c@{\hspace{5mm}}ccc@{\hspace{5mm}}c}
\hline
\hline
obs & fit & $1$ & $1/N_c$ & $1/N_c^2$ & $p$-value \\[0.3ex] 
\hline\\[-2.5ex]
$\hat{B}_K$  & l, $N_c \geq 3$ & 0.802(17) & -0.03(10) & --- & 0.24 \\
             & l, $N_c \geq 4$ & 0.808(27) & -0.07(16) & --- & 0.14 \\
             & q, $N_c \geq 3$ & 0.788(79) &  0.12(78) & -0.3(1.8) & 0.12 \\[0.3ex]
\hline\\[-2.5ex]
$A^+$        & l, $N_c \geq 3$ & 0.956(20) & -0.89(11) & --- & 0.10 \\
             & l, $N_c \geq 4$ & 0.981(18) & -1.05(11) & --- & 0.39 \\[0.3ex]
\hline\\[-2.5ex]
$A^-$        & l, $N_c \geq 3$ & 0.984(28) &  1.77(17) & --- & 0.21 \\
             & l, $N_c \geq 4$ & 0.996(39) &  1.69(24) & --- & 0.14 \\[0.3ex]
 \hline
\hline
\end{tabular}
\caption{Fit parameters of $A^\sigma$ assuming a linear (l) or quadratic (q) dependence, and various fit ranges.
The order at which each coefficient enters in the polynomial ansatz in powers of
$1/N_c$ is indicated, alongside with the $p$-value for each fit.}
\label{tab:fits}
\end{center}
\end{table}

\begin{figure}[!t]
 \begin{center}
 \includegraphics[scale=0.68]{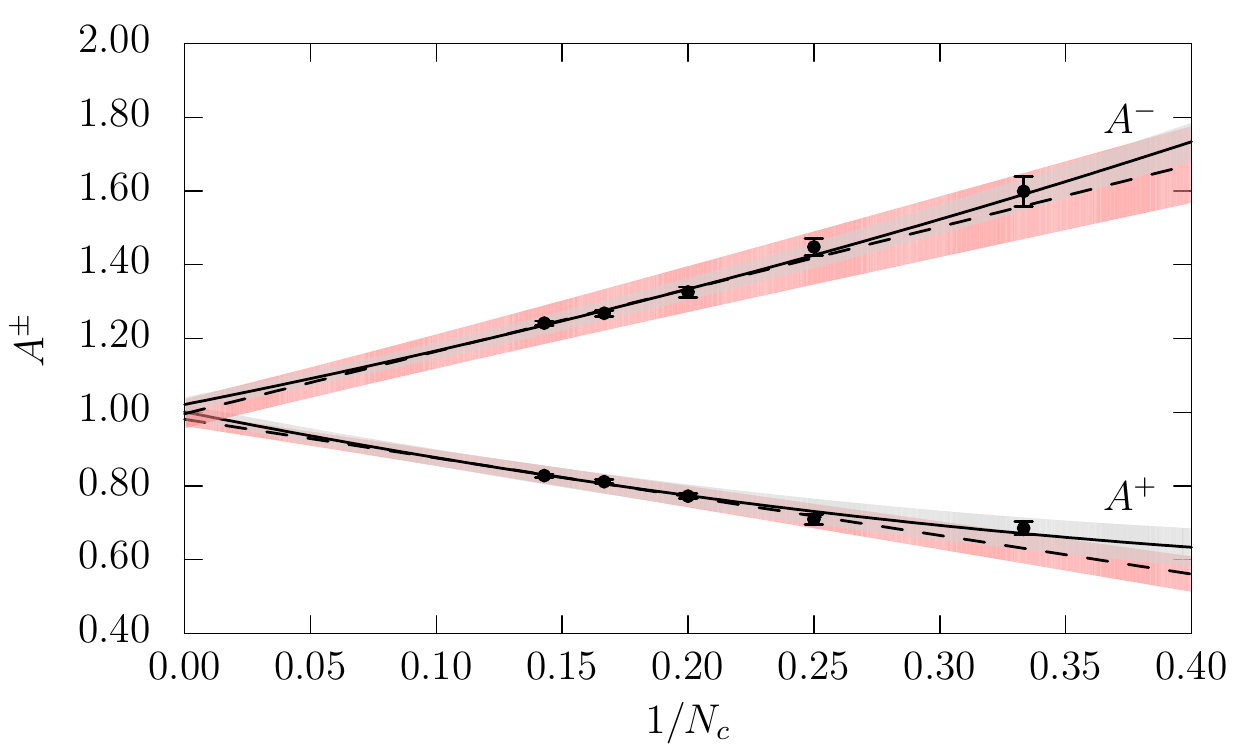}  
\caption{\label{fig:apm}  $A^\pm$ versus $1/N_c$. 
The grey bands (solid lines) are obtained from the results of the fits to $1/2 (A^- \pm A^+)$ in eqs.~(\ref{eq:fit}); the red bands (dashed lines) are linear fits including $N_c=4-7$ from Table~\ref{tab:fits}.
}
\end{center}
\end{figure}

\begin{figure}[h]
 \begin{center}
 \includegraphics[scale=0.68]{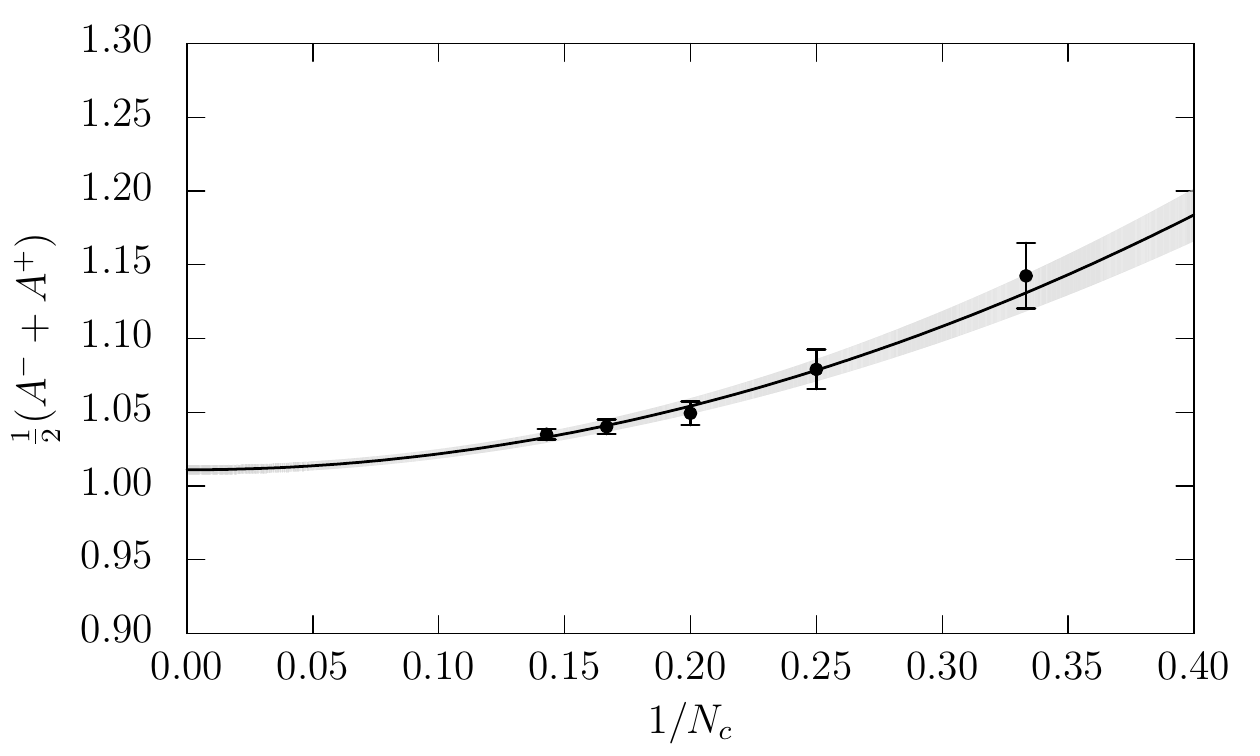} 
 \includegraphics[scale=0.68]{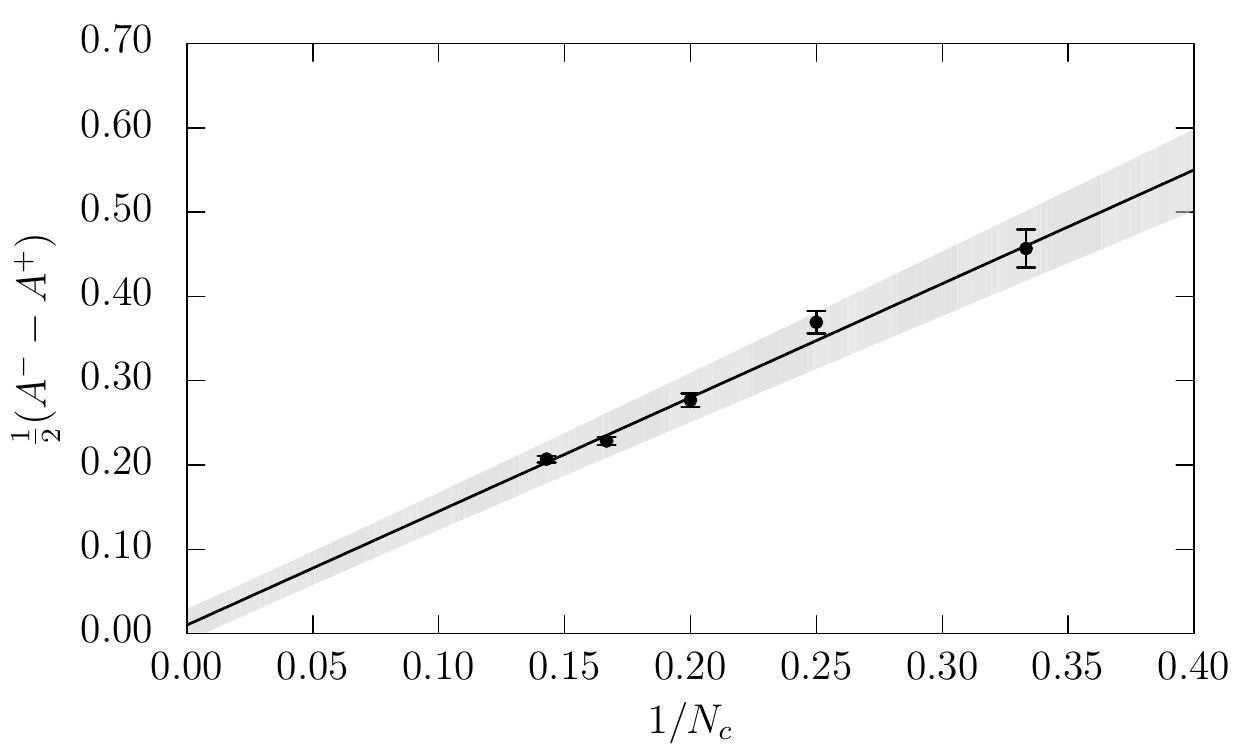} 
 \caption{\label{fig:comb}   ${A^-\pm A^+\over 2}$ versus $1/N_c$.
The bands (solid lines) are quadratic and linear fits in $1/N_c$, respectively.}
\end{center}
\end{figure}

The smallness of $1/N_c$ corrections in $\hat{B}_K$ is related to the RGI normalization of this quantity, $\hat c^+(a^{-1})$:
the significant $N_c$-dependence of $R^+$ (see Table~\ref{tab:sim}) is cancelled
to a large extent by the RGI Wilson coefficient $\hat k^+$ (see Table~\ref{tab:renorm}).
In contrast,  the total $K\to\pi$ amplitudes  show  very significant subleading $1/N_c$ corrections, as shown in Fig.~\ref{fig:apm}.
In the Figure we present our data for $A^\pm$ obtained from the ratios $R^\pm$ of eq.~(\ref{eq:bareratios}) and the results of a linear (dashed lines) and quadratic (solid lines)
fit to the data. The parameters of the linear fit for $A^+$ and $A^-$ are shown in the fourth and fifth (sixth and seventh) lines of Table~\ref{tab:fits}, respectively. 
We can see from the Figure that the corrections at $N_c=3$ are naturally  $\sim 30\%$ and that they are strongly anti-correlated in $A^\pm$. 
For the quadratic fit, and in order to clarify further this correlation, we have considered  the combinations ${1\over 2}(A^-\pm A^+)$; the results are shown in Fig.~\ref{fig:comb}. 
The curves correspond to the following best fits:
\begin{gather}
\begin{split}
{A^-+A^+\over 2} &= 1.01(3) + {1.08(11) \over N^2_c} \quad (p{\rm -value} =0.81), \\
{A^--A^+\over 2} &= 0.01(2) + {1.35(11) \over N_c}   \quad (p{\rm -value} =0.12). 
\end{split}
\label{eq:fit}
\end{gather}
The subleading $1/N_c$ effects seem to cancel in the first combination, while they are the only visible corrections in the second one.  
The parameters of the quadratic fit in Fig.~\ref{fig:apm} are obtained from the results of eq.~(\ref{eq:fit}).

We have not included any systematic error in these results. There are two obvious sources: finite lattice spacing and the quenched approximation. Although 
it is impossible to quantify those errors, we do not expect them to be larger that those observed at $N_c=3$, where they have been studied.
We have already commented above on the expected size of $O(a^2)$ discretization effects,
based on the results of~\cite{Constantinou:2010qv}.
Concerning the quenching error, it is well-known that $\hat B_K$ is remarkably insensitive
to the number of dynamical quark flavours, cf. \cite{Aoki:2013ldr} and benchmark quenched studies
\cite{qBK}; we thus expect a small effect in $A^+$.
The pioneering large-$N_c$ study of dynamical QCD in~\cite{DeGrand:2016pur} shows that an extension
of our work to take into account unquenching effects is feasible.

\section{Conclusions}
\label{sec:conclusions}

We have presented the first computation on the lattice of  the $1/N_c$ corrections to the $\Delta S=1$ amplitudes $K-\pi$ in the GIM and SU(3) limit $m_c=m_u=m_s=m_d$. The size and sign of $1/N_c$ 
corrections are relevant to give a solid physical basis to the observation made in \cite{Boyle:2012ys} that suggests that the $\Delta I=1/2$ rule might originate in a near cancellation of two contributions to the $K\rightarrow (\pi\pi)_{I=2}$ amplitude, that add up in the $I=0$ channel.  The observed cancellation can be  traced to large and anti-correlated $1/N_c$ corrections in the two isospin amplitudes.  We have quantified the subleading $1/N_c$ dependence of the simpler $K-\pi$ amplitudes, $A^\pm$,  that are closely related to  the $K-\pi\pi$ ones in the degenerate light quark limit, $m_s = m_d$.  Our results show that the subleading $1/N_c$ corrections in $A^\sigma$ are large and consistent with being equal and 
opposite in sign for $A^+$ and $A^-$, supporting the observation in \cite{Boyle:2012ys}. However, the size of these corrections is natural, i.e. ${\mathcal O}(1)/N_c$ and not large enough to explain the $\Delta I=1/2$ rule, although we have argued that larger $1/N_c$ corrections could be present at the physical point, $m_s \gg m_d$, suggested by a large chiral log. We have also studied the subleading $N_c$ corrections to $\hat{B}_K$ and found that they are significantly smaller than those in the closely related amplitude $A^+$, because of the different normalization. This shows that a value of $\hat{B}_K$ close to the $N_c \rightarrow \infty$ value is consistent with large $1/N_c$ corrections in the $\Delta S=1$ amplitudes.

\begin{acknowledgments}

We warmly thank  C.~Pica for providing us with a $SU(N_c)$ lattice code. 
This work was partially supported by grants FPA2012-31686, FPA2014-57816-P, FPA2015-68541-P (MINECO/FEDER), PROMETEOII/2014/050, MINECO's ``Centro de Excelencia Severo Ochoa'' Programme under grants SEV-2012-0249 and SEV-2014-0398, 
and the European projects 
H2020-MSCA-ITN-2015//674896-ELUSIVES and H2020-MSCA-RISE-2015. 

\end{acknowledgments}


\vspace{0.5cm}
\begin{thebibliography}{99}

\bibitem{Bai:2015nea}
  Z.~Bai {\it et al.} [RBC and UKQCD Collaborations],
  Phys.\ Rev.\ Lett.\  {\bf 115} (2015) no.21,  212001.


\bibitem{Aoki:2013ldr}
S.~Aoki {\it et al.},
  Eur.\ Phys.\ J.\ C {\bf 74} (2014) 2890;
S.~Aoki {\it et al.},
  arXiv:1607.00299 [hep-lat].

\bibitem{BK}
S.~D\"urr {\it et al.},
  Phys.\ Lett.\ B {\bf 705} (2011) 477;
J.~Laiho and R.S.~Van de Water,
  PoS LATTICE {\bf 2011} (2011) 293;
T.~Blum {\it et al.} [RBC and UKQCD Collaborations],
  Phys.\ Rev.\ D {\bf 93} (2016) no.7,  074505;
B.J.~Choi {\it et al.} [SWME Collaboration],
  Phys.\ Rev.\ D {\bf 93} (2016) no.1,  014511;
N.~Carrasco {\it et al.} [ETM Collaboration],
  Phys.\ Rev.\ D {\bf 92} (2015) no.3,  034516.
  

\bibitem{'tHooft:1973jz}
  G.~'t~Hooft,
  Nucl.\ Phys.\ B {\bf 72} (1974) 461.


\bibitem{Buras:2014maa}
  A.J.~Buras, J.M.~G\'erard and W.A.~Bardeen,
  Eur.\ Phys.\ J.\ C {\bf 74} (2014) 2871.

 \bibitem{Pich:1995qp}
  A.~Pich and E.~de~Rafael,
  Phys.\ Lett.\ B {\bf 374} (1996) 186.
  
\bibitem{Peris:2000sw}
  S.~Peris and E.~de~Rafael,
  Phys.\ Lett.\ B {\bf 490} (2000) 213.
  

\bibitem{Hambye:2003cy}
  T.~Hambye, S.~Peris and E.~de~Rafael,
  JHEP {\bf 0305} (2003) 027.


\bibitem{Cirigliano:2011ny}
  V.~Cirigliano, G.~Ecker, H.~Neufeld, A.~Pich and J.~Portol\'es,
  Rev.\ Mod.\ Phys.\  {\bf 84} (2012) 399.


\bibitem{Boyle:2012ys}
  P.A.~Boyle {\it et al.} [RBC and UKQCD Collaborations],
  Phys.\ Rev.\ Lett.\  {\bf 110} (2013) no.15,  152001.
  

\bibitem{Blum:2015ywa}
  T.~Blum {\it et al.},
  Phys.\ Rev.\ D {\bf 91} (2015) no.7,  074502.

\bibitem{Carrasco:2013jda}
  N.~Carrasco {\it et al.} [ETM Collaboration],
  Phys.\ Lett.\ B {\bf 736} (2014) 174.


\bibitem{di32}
J.F.~Donoghue, E.~Golowich and B.R.~Holstein,
  Phys.\ Lett.\ B {\bf 119} (1982) 412;
J.~Bijnens, H.~Sonoda and M.B.~Wise,
  Phys.\ Rev.\ Lett.\  {\bf 53} (1984) 2367.
  

\bibitem{Giusti:2004an}
  L.~Giusti {\it et al.},
  JHEP {\bf 0411} (2004) 016.


\bibitem{Giusti:2006mh}
  L.~Giusti {\it et al.},
  Phys.\ Rev.\ Lett.\  {\bf 98} (2007) 082003.


\bibitem{Shifman:1975tn}
  M.A.~Shifman, A.I.~Vainshtein and V.I.~Zakharov,
  Nucl.\ Phys.\ B {\bf 120} (1977) 316.


\bibitem{Endress:2014ppa}
  E.~Endress and C.~Pena,
  Phys.\ Rev.\ D {\bf 90} (2014) 094504.


\bibitem{b0}
D.J.~Gross and F.~Wilczek,
  Phys.\ Rev.\ Lett.\  {\bf 30} (1973) 1343;
H.D.~Politzer,
  Phys.\ Rev.\ Lett.\  {\bf 30} (1973) 1346.


\bibitem{b1}
W.E.~Caswell,
  Phys.\ Rev.\ Lett.\  {\bf 33} (1974) 244;
D.R.T.~Jones,
  Nucl.\ Phys.\ B {\bf 75} (1974) 531;
E.~Egorian and O.V.~Tarasov,
  Teor.\ Mat.\ Fiz.\  {\bf 41} (1979) 26
   [Theor.\ Math.\ Phys.\  {\bf 41} (1979) 863].

  
\bibitem{load}
M.K.~Gaillard and B.W.~Lee,
  Phys.\ Rev.\ Lett.\  {\bf 33} (1974) 108;
G.~Altarelli and L.~Maiani,
  Phys.\ Lett.\ B {\bf 52} (1974) 351.

\bibitem{nprunning}
M.~Guagnelli {\it et al.} [ALPHA Collaboration],
  JHEP {\bf 0603} (2006) 088;
P.~Dimopoulos {\it et al.} [ALPHA Collaboration],
  JHEP {\bf 0805} (2008) 065.


\bibitem{nload}
M.~Ciuchini {\it et al.},
  Nucl.\ Phys.\ B {\bf 523} (1998) 501;
A.J.~Buras, M.~Misiak and J.~Urban,
  Nucl.\ Phys.\ B {\bf 586} (2000) 397.


\bibitem{Golterman:1997wb}
  M.F.L.~Golterman and K.C.~Leung,
  Phys.\ Rev.\ D {\bf 56} (1997) 2950.


\bibitem{Truong:1987hn}
  T.~N.~Truong,
  Phys.\ Lett.\ B {\bf 207} (1988) 495.
  doi:10.1016/0370-2693(88)90690-9


\bibitem{Isgur:1989js}
  N.~Isgur, K.~Maltman, J.~D.~Weinstein and T.~Barnes,
  Phys.\ Rev.\ Lett.\  {\bf 64} (1990) 161.
  doi:10.1103/PhysRevLett.64.161


\bibitem{Kambor:1991ah}
  J.~Kambor, J.H.~Missimer and D.~Wyler,
  Phys.\ Lett.\ B {\bf 261} (1991) 496.
  doi:10.1016/0370-2693(91)90463-Z

\bibitem{Pallante:2000hk}
  E.~Pallante and A.~Pich,
  Nucl.\ Phys.\ B {\bf 592} (2001) 294
  doi:10.1016/S0550-3213(00)00601-5
  [hep-ph/0007208].

\bibitem{Hernandez:2006kz}
  P.~Hern\'andez and M.~Laine,
  JHEP {\bf 0610} (2006) 069.

\bibitem{DelDebbio:2008zf}
  L.~Del Debbio, A.~Patella and C.~Pica,
  Phys.\ Rev.\ D {\bf 81} (2010) 094503.


\bibitem{pica}
C.~Pica, private communication. 


\bibitem{Bali:2013kia}
  G.S.~Bali {\it et al.},
  JHEP {\bf 1306} (2013) 071.


\bibitem{tmqcd}
R.~Frezzotti {\it et al.} [ALPHA Collaboration],
  JHEP {\bf 0108} (2001) 058;
R.~Frezzotti and G.C.~Rossi,
  JHEP {\bf 0408} (2004) 007.


\bibitem{Frezzotti:2004wz}
  R.~Frezzotti and G.C.~Rossi,
  JHEP {\bf 0410} (2004) 070
  doi:10.1088/1126-6708/2004/10/070
  [hep-lat/0407002].
  

\bibitem{etmpert}
M.~Constantinou {\it et al.},
  Phys.\ Rev.\ D {\bf 83} (2011) 074503;
 C.~Alexandrou, M.~Constantinou, T.~Korzec, H.~Panagopoulos and F.~Stylianou,
  Phys.\ Rev.\ D {\bf 86} (2012) 014505.


\bibitem{Allton:2008ty}
  C.~Allton, M.~Teper and A.~Trivini,
  JHEP {\bf 0807} (2008) 021.


\bibitem{Constantinou:2010qv}
  M.~Constantinou {\it et al.} [ETM Collaboration],
  Phys.\ Rev.\ D {\bf 83} (2011) 014505.


\bibitem{qBK}
  S.~Aoki {\it et al.} [JLQCD Collaboration],
  Phys.\ Rev.\ Lett.\  {\bf 81} (1998) 1778;
  S.~Aoki {\it et al.} [JLQCD Collaboration],
  Phys.\ Rev.\ Lett.\  {\bf 80} (1998) 5271;
  P.~Dimopoulos {\it et al.} [ALPHA Collaboration],
  Nucl.\ Phys.\ B {\bf 749} (2006) 69;
  P.~Dimopoulos {\it et al.} [ALPHA Collaboration],
  Nucl.\ Phys.\ B {\bf 776} (2007) 258.


\bibitem{DeGrand:2016pur}
  T.~DeGrand and Y.~Liu,
  Phys.\ Rev.\ D {\bf 94} (2016) no.3,  034506.


\end{thebibliography}
\end{document}